\documentclass[]{jfm_arXiv}
\usepackage{graphicx}
\usepackage{newtxtext}
\usepackage{newtxmath}
\usepackage{natbib}
\usepackage{subfig}
\usepackage{hyperref}
\hypersetup{
    colorlinks = true,
    urlcolor   = blue,
    citecolor  = blue,
}

\newcommand{\RomanNumeralCaps}[1]
\linenumbers
\title{A Robust Physics-based Method to Filter Coherent Wavepackets from High-speed Schlieren Images}
\author{Chitrarth Prasad
  \corresp{\email{prasad.141@osu.edu}}
  \and Datta V. Gaitonde}

\affiliation{Department of Mechanical and Aerospace Engineering, The Ohio State University, OH 43210}

\usepackage{xargs}
\usepackage[colorinlistoftodos,prependcaption,textsize=tiny]{todonotes}
\newcommandx{\unsure}[2][1=]{\todo[linecolor=red,backgroundcolor=red!25,bordercolor=red,#1]{#2}}
\newcommandx{\needswork}[2][1=]{\todo[linecolor=blue,backgroundcolor=blue!25,bordercolor=blue,#1]{#2}}
\newcommandx{\info}[2][1=]{\todo[linecolor=OliveGreen,backgroundcolor=OliveGreen!25,bordercolor=OliveGreen,#1]{#2}}
\newcommandx{\improvement}[2][1=]{\todo[linecolor=Plum,backgroundcolor=Plum!25,bordercolor=Plum,#1]{#2}}
\newcommandx{\thiswillnotshow}[2][1=]{\todo[disable,#1]{#2}}
\usepackage{ulem}
\usepackage{soul}

\begin{document}
\maketitle

\begin{abstract}
A complete understanding of jet dynamics is greatly enabled by accurate separation of the acoustically efficient wavepackets from their higher-energy convecting turbulent counterparts. 
The filtering operation may be achieved by one of several methods.
For example, signal processing techniques based on wave speed with user-defined, operating condition dependent thresholds may be applied in the near-acoustic field.
Recent developments using Momentum Potential Theory (MPT) have successfully isolated the acoustic component in all regions of the jet, to better understand the dynamics as well as to develop wavepacket models.
MPT is however a data-intensive method since 
the inherent Poisson equation solution requires fluctuation quantities in the entire flowfield; as such, it has to date been applied only to numerically obtained data.
This work develops an approach to extend its application to extract coherent wavepacket data from high-speed schlieren images.
The procedure maps pixel intensities from the schlieren to a scaled surrogate for the density gradient integrated along the line of sight.
The linear relation between the irrotational scalar MPT potential and time-derivatives of density fluctuations is then exploited to perform the filtering.
The effectiveness of the procedure is demonstrated using experimental as well as simulated schlieren images representing a wide range of
imperfectly-expanded free and impinging jet configurations. 
When combined with Spectral Proper Orthogonal Decomposition, the method yields modes that accurately capture (i) the Mach wave radiation from a military-style jet, (ii) the mode shapes of the feedback tones in an impinging jet, and (iii) the screech signature in twin rectangular jets,  without recourse to user adjusted parameters.
This technique has the potential to greatly expand the use of high-speed diagnostics and provide real-time monitoring of the acoustic content of the jet in the nearfield, with feedback control implications.
Additionally, although the present study focuses on jets, the general nature of the approach 
allows a straightforward application to other flows, such as cavity flow-acoustic interactions, among others.


\end{abstract}



\section{Introduction}
\label{sec:intro}
Acoustic waves play a significant role in jet dynamics.
In addition to determining the far-field noise, they also provide feedback paths to trigger instabilities.
Therefore, a crucial benefit in understanding different aspects of jet dynamics and exploring flow control accrues if the jet acoustic content can be filtered from the rest of the turbulent flowfield. 
This is not a trivial process in the turbulent core or the near-acoustic field, since the acoustic energy forms only a tiny fraction of the total energy contained in a jet.

Nevertheless, the acoustic component in the jet near-acoustic field exhibits certain unique characteristics~\citep{jordan2013wave} that facilitate the capture of the near-field acoustic footprint with signal processing techniques. 
\cite{arndt1997proper} performed a proper orthogonal decomposition (POD) of pressure fluctuations measured along the outer edge of a jet shear layer and argued that near-field pressure spectra in a subsonic jet can be segregated into a low-frequency hydrodynamic and a high-frequency acoustic regime.
This idea was expanded to incorporate phase speeds by \cite{tinney2008near}, who applied a wavenumber-frequency bandpass filter on near-field pressure measurements, obtained from a similar line array, to separate the convecting and radiating pressure components in coaxial subsonic jets.
Selective frequency filtering was extended to supersonic jets by \cite{kuo2013experimental}, who applied an empirical modal decomposition to separate the near-field pressure spectra into intrinsic mode functions, which were then combined to represent the contribution of hydrodynamic and acoustic components.
\cite{grizzi2012wavelet} leveraged the intermittent nature of the near-field pressure fluctuations to develop a wavelet-based approach to reconstruct the time series of hydrodynamic and acoustic components from a two-dimensional array of pressure measurements. 
The method was further advanced by \cite{mancinelli2017wavelet} to rely on minimal data 
by incorporating probability density functions, and cross-correlations with far-field measurements to extract the acoustic component from each pressure probe.

Although the above-mentioned methods have been successful, they generally require specification of thresholds that depend on the jet operating condition.
An alternate method that has proven effective in extracting the acoustic component is Momentum Potential Theory (MPT)~\citep{doak1989momentum}.
As summarized in $\S$~\ref{sec:MPT}, an acoustic component, defined as the irrotational-isentropic part of the ``momentum density'', $\rho\boldsymbol{u}$ 
may be extracted through a Poisson equation with a source term that depends on primitive variable fluctuations.
The other two components, hydrodynamic (vortical) and thermal (irrotational-isobaric), may also be suitably extracted.
The formulation is exact regardless of the size of turbulent fluctuations or the presence of other complicating factors such as shock cells, because it leverages the linearity of the mass conservation equation in $\rho\boldsymbol{u}$.

MPT-derived components offer significant insights into jet dynamics, which are difficult to obtain with the more commonly used primitive variables. 
Recent efforts have successfully used MPT to understand exchange mechanisms by which the chaotic turbulence energy in a free jet is converted into sound~\citep{unnikrishnan2016acoustic,unnikrishnan2018transfer}, to predict far-field noise based on a simple linear wave propagator from the edge of the turbulent jet and superior wavepacket models~\citep{unnikrishnan2019acoustically}, to investigate noise reduction mechanisms of fluid inserts in over-expanded heated military jets~\citep{prasad2019effect,prasad2020study,prasad2021steady} and to analyse receptivity and generation processes in an under-expanded impinging jet~\citep{prasad2021exchange}.
These examples highlight the versatile nature of the MPT procedure which poses no restrictions on the complexity of the problem, that can otherwise present an additional challenge to traditional signal processing techniques. 

Despite its effectiveness, the application of Doak's MPT to date has been mostly predicated on the availability of rich numerical databases generated with substantial computational resources.
In fact, a complete application of MPT, as shown in $\S$\ref{sec:MPT1}, requires a simultaneous space-time description of momentum density, pressure and density field, which is currently beyond the scope of experimental techniques.
Although the computational cost is significantly relaxed when the MPT-derived components are obtained as instability modes about the mean flow as in \citet{prasad2021extraction}, there is an inherent advantage in extending this procedure to experimental measurements, which generally contain significantly longer time series data than numerical simulations.
This facilitates better statistical convergence and excellent resolution of low-frequency phenomena, which are very computationally demanding to capture. 
Measurements are, however, often limited by their sampling resolution and the number of quantities that can be measured simultaneously. 

The present work extends crucial elements of the MPT procedure to experimental measurements by suitably interpreting and processing time-resolved high-speed schlieren data, whose resolution and fidelity has seen rapid improvement in recent years.
Schlieren imaging is a widely used flow visualization technique that can provide an accurate space-time description of the density gradient field, and therefore forms the basis of the present study. 
We show that the original MPT procedure ($\S$\ref{sec:MPT2}) can be modified to incorporate the pixel intensities extracted from schlieren images. 
However, since schlieren images provide only the density gradients in the flow, the decomposition is partial and essentially extracts the combination of irrotational isentropic and isobaric components.
However, the isobaric component is not critical outside the turbulent core, and thus, the resulting wavepacket structures from schlieren-derived MPT accurately capture the observed near-field acoustic characteristics.
This is demonstrated for  
both free and impinging jets in $\S$\ref{sec:TestCases} using configurations selected based on their relevance to practical jet noise problems and the availability of high-quality large eddy simulations (LES) and experiments.
The numerical test cases serve as truth models; for these, 
schlieren images are constructed from the 3D LES flow-field using the procedure laid out by \cite{yates1993images}.
The procedure has been demonstrated to provide accurate one-on-one comparisons between experiments and computations for general 3D flow-fields.
The resulting wavepackets from the schlieren data are validated with those obtained from the full LES data using Spectral Proper Orthogonal Decomposition (SPOD)~\citep{towne2018spectral}. 
For the experimental test case, the SPOD modes are validated using available near-field measurements.
A summary of the present findings is presented in $\S$\ref{sec:Conclusion}.


\section {Method \label{sec:MPT}}
This section provides pertinent details of MPT and its implementation to both LES and schlieren data. 
\subsection {General Formulation \label{sec:MPT1}} 
MPT comprises a Helmholtz decomposition of the ``momentum-density'' $(\rho \boldsymbol{u})$ flow-field into its solenoidal and irrotational components:
\begin{equation} \label{Doak1}
\rho \boldsymbol u = \bar{\boldsymbol B} + \boldsymbol B' - \nabla \psi', \hspace{0.2in} \nabla \cdot\bar{\boldsymbol B}=0, \hspace{0.2in }\nabla \cdot\boldsymbol B'=0.
\end{equation}
where
$\bar{\boldsymbol B}$ is the mean solenoidal component, $\boldsymbol  B'$ is the fluctuating solenoidal component and $\psi'$ is the fluctuating irrotational scalar potential. 
Equations~\ref{Doak1} 
when substituted into the continuity equation yield,
\begin{equation}\label{Doak_irr}
\nabla^2 \psi'=\frac{\partial \rho'}{\partial t}.
\end{equation}
The irrotational scalar potential can be further split into its acoustic and thermal components, 
    $\psi'=\psi'_a + \psi'_t$, 
where the acoustic component is described by 
\begin{equation} \label{Doak_irr2}
\nabla^2 \psi_a'=\frac{1}{c^2}\frac{\partial p'}{\partial t}.
\end{equation}

The procedure~\citep{unnikrishnan2016acoustic}  requires the solution of the two 
Poisson equations (\ref{Doak_irr} and~\ref{Doak_irr2}) 
to obtain $\psi'$, $\psi_a'$ and $\psi_t'$. 
Since the outer boundaries in a numerical solution are often far away from the jet, these are taken as purely acoustic; this results in $(\rho \boldsymbol{u})'=-\nabla \psi' = -\nabla \psi'_a$ which is integrated along the boundaries to provide a Dirichlet boundary condition to the Poisson solver.
Once $\psi'$ is known, the fluctuating solenoidal component is obtained from eqn.~\ref{Doak1}.
This solution procedure has been extensively validated for a wide array of flow configurations including both free and impinging jets at on- and off-design conditions~\citep{unnikrishnan2018transfer,prasad2020study,prasad2021exchange}. 

\subsection{Application to Schlieren Data \label{sec:MPT2}}
A key feature in MPT-based investigations is the use of the streamwise gradient of the acoustic potential, $-\partial \psi'_a/\partial x$, as a representation of the acoustic content in the jet shear layer. 
In free jets, $-\partial \psi'_a/\partial x$ exhibits a highly coherent wavepacket structure that scales with pressure fluctuations away from the jet shear layer, whereas in impinging jets, $-\partial \psi'_a/\partial x$ accurately captures the upstream radiating component of the feedback loop. 
This information is leveraged when applying the MPT procedure to schlieren data in the following manner.

\begin{figure}
    \centering
    \subfloat[\label{fig:rays}]{\includegraphics[width=0.5\textwidth, trim={20 180 20 60},clip]{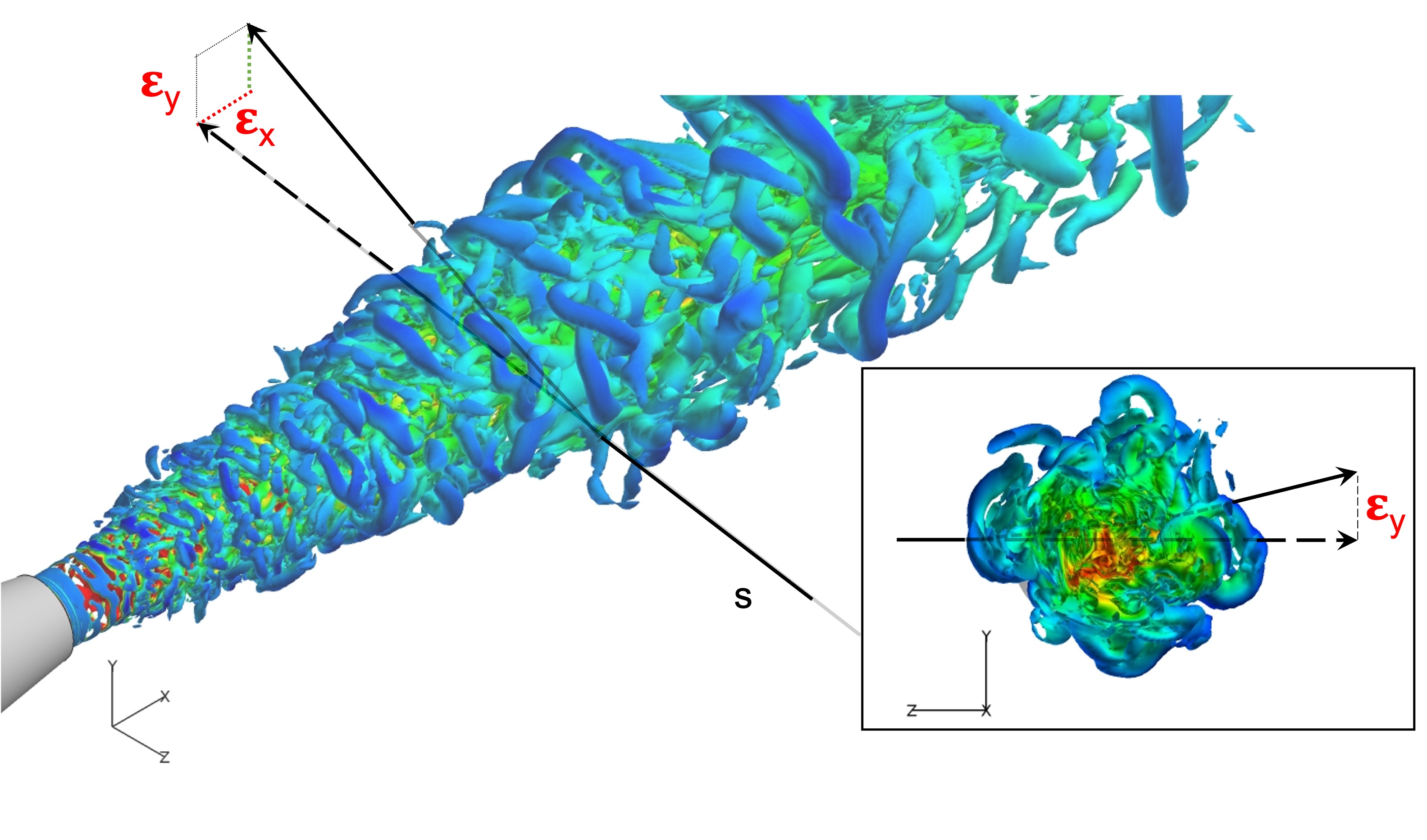}}
    \subfloat[\label{fig:TestCase1}]{\includegraphics[width=0.5\textwidth,trim=0 0 20 0, clip]{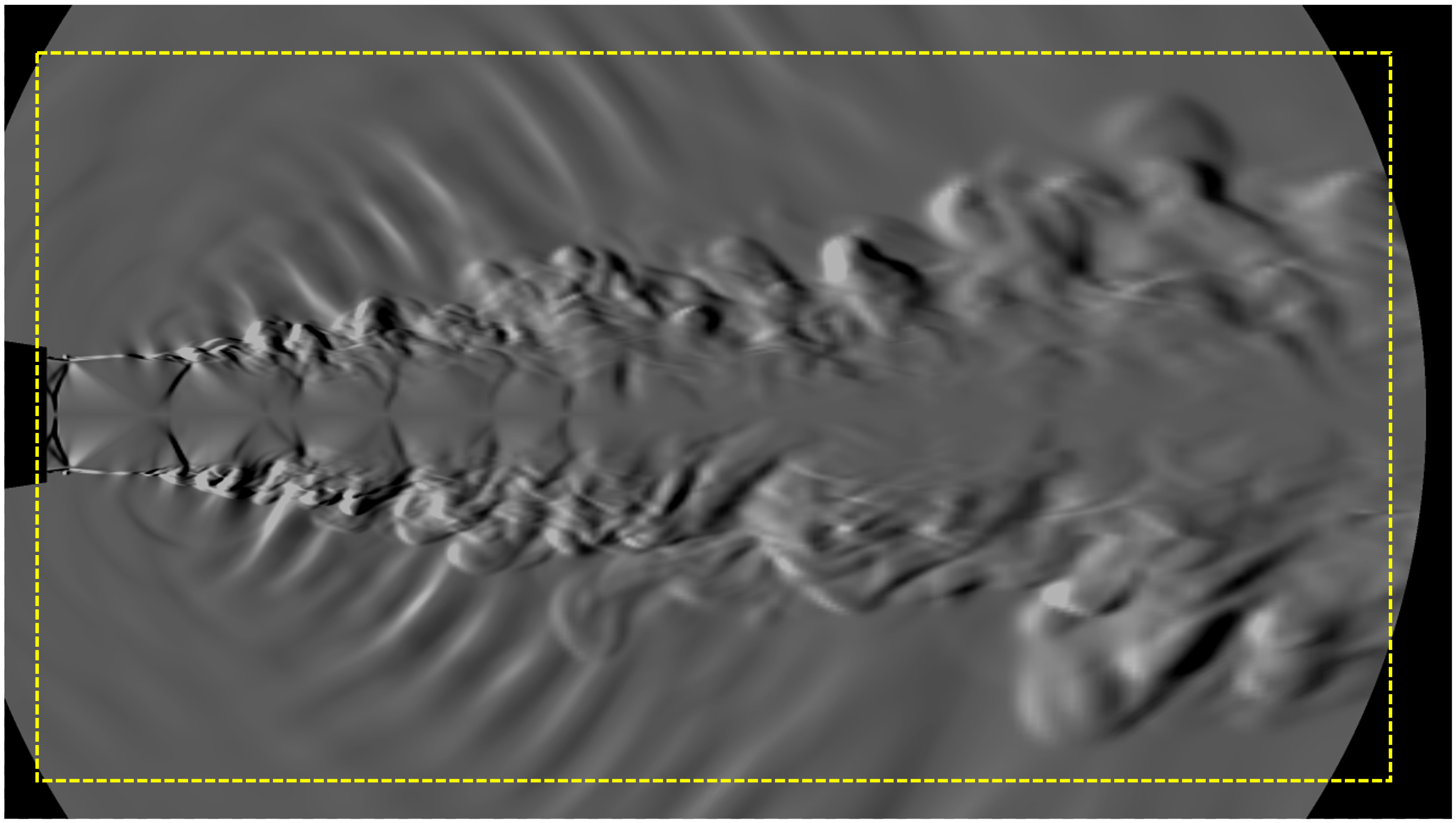}}
    \caption{(a) Schematic showing the deflection of a light ray cutting through a jet, (b) Instantaneous schlieren image for the heated over-expanded Mach 1.36 GE-F404 test case. }
    
\end{figure}
Schlieren images are created by passing collimated light through a 3-D flow-field, where they undergo angular deflection. 
The rays exiting the flow are then focused at a point where a portion of the light is blocked by a knife-edge before refocusing the remaining light onto an image plane. 
The image essentially captures the deflection of light in the directional perpendicular to the knife edge. 
Figure~\ref{fig:rays} shows a schematic of a light ray passing through a jet.
The total angular deflection of the light ray can be obtained by integrating functions of the refractive index ($n$) along lines of sight:
\begin{equation}\label{eqn:deflection}
    \epsilon_x =  \int \frac{1}{n} \frac{\partial n}{\partial x} \text{ds}, \hspace{0.1in} \epsilon_y =  \int \frac{1}{n} \frac{\partial n}{\partial y} \text{ds},
\end{equation}
where $\epsilon_x$ and $\epsilon_y$ are the angular deflections in the $x-$ and $y-$ direction respectively, and $s$ denotes the path travelled by the light ray.
For ideal and non-reacting gases, the refractive index $n$ is simply
\begin{equation}\label{eqn:GD}
    n = 1 + \kappa \rho,
\end{equation}
where $\kappa$ is the Gladstone–Dale constant ($\kappa= 2.23 \times 10^{-4} \text{m}^3/\text{kg}$ for air).
Following this relation, the total angular deflection in the streamwise direction may be rewritten as
\begin{equation} \label{eqn:deflection}
    \epsilon_x =  \int \frac{\partial}{\partial x}\log(1 + \kappa \rho) \text{ds} \approx \int \kappa \frac{\partial \rho}{\partial x} \text{ds}.
\end{equation}
The pixel intensities from a schlieren image therefore, provide a scaled surrogate for the flow density gradient integrated along the line of sight. 
Although the present method is valid for gradients in any direction, 
for reasons stated earlier,
we only consider a vertical orientation of the knife-edge throughout this investigation \textit{i.e.,} all the pixel intensities considered are proportional to streamwise density gradients.

Due to the linearity of in $\psi'$, eqn.~{\color{red}\ref{Doak_irr}} can be modified to directly obtain
\begin{equation} \label{eqn:Doakxgrad}
\nabla^2 \left( \frac{\partial \psi'}{\partial x}\right) = \frac{\partial}{\partial t}\left(\frac{\partial \rho}{\partial x}\right)'. 
\end{equation}
or,
\begin{equation} \label{eqn:DoakSch}
\nabla^2 \left( \Theta' \right) = \frac{\partial \sigma'}{\partial t}, 
\end{equation}
where $\sigma$ is the image intensity obtained from schlieren (proportional to $\partial \rho/\partial x$) and $\Theta$ is the resulting image intensity which is representative of $\partial \psi/\partial x$, 
the gradients of the irrotational MPT component integrated along the line of sight of the schlieren image.
As mentioned earlier, the streamwise gradient of $\psi'_a$ best describes the acoustic content in the jet shear layer.
The resulting $\Theta'$ from the schlieren images however, contains contributions due to both the acoustic and thermal components.
Previous numerical investigations have shown that the thermal component ($-\partial \psi'_t/\partial x$), similar to its acoustic counterpart ($-\partial \psi'_a/\partial x$), exhibits a coherent wavepacket-like structure, which
is non-radiating even for highly heated supersonic jets~\citep{prasad2021steady}.
This enables the primary noise radiating mechanisms to be extracted from the schlieren-derived $\Theta'$, as demonstrated in the test cases below.



\section{Test Cases}\label{sec:TestCases}
\subsection{Test Case 1: Heated Over-expanded Jet}
A model-scale GE-F404 military-style, faceted nozzle ($D=0.0254\text{m}$) is considered first at 
Mach~$1.36$, a nozzle pressure ratio of $3.0$ and a total temperature ratio of $2.5$. 
The LES database 
has been previously validated extensively against experimental measurements~\citep{prasad2019effect,prasad2020study}.
A total of $1{,}500$ Portable Network Graphic 
images ($720p$) are first generated from the 3D LES snapshots sampled at $200$kHz by integrating the expression for $\epsilon_x$ in eqn.~\ref{eqn:deflection} following the procedure of \citet{yates1993images}. 
The 
schlieren domain extends up to $10D$ downstream of the nozzle exit and $2.5D$ in the radial direction to represent a practical field of view obtainable with experiments.
In contrast, the LES domain extends up to $65D$ downstream and $25D$-$40D$ radially depending on the streamwise location.

Figure~\ref{fig:TestCase1} shows an instantaneous schlieren image at an arbitrary time instant.
The schlieren image clearly shows the turbulent structures in the jet shear layer 
and the shock-cells due to the over-expanded nature of the jet.
In addition, the strong downstream Mach wave radiation signature characteristic of heated supersonic jets is also captured by the simulated schlieren 
To quantify the source term in eqn.~\ref{eqn:DoakSch}, the pixel intensities in the schlieren image are converted to double precision floating-point decimals using MATLAB. 
Even though this yields a space-time description of the source term, a simultaneous measurement of $\rho \boldsymbol{u}$ is not.
This constraint, along with the smaller extent of the schlieren domain relative to the computations, prohibits a specification of a purely acoustic boundary condition for the Poisson solver, unlike for application to LES data.

This limitation is overcome by incorporating a sponge zone to gradually damp the acoustic waves to zero as they reach the outer boundaries; this results in a $\Theta'=0$ boundary condition for the Poisson solver.
Figure~\ref{fig:TestCase1} highlights the sponge zone as the region between the outer boundaries and the yellow box.
Equation~\ref{eqn:DoakSch} is then discretized to a second-order accuracy and solved using the BiCGSTAB algorithm 
to obtain a corresponding $\Theta'$ for each snapshot.

The resulting $\Theta'$ snapshots are subsequently processed with SPOD. 
The Fourier transform is performed by arranging the snapshots into blocks of $256$ with a $50\%$ overlap. 
A Hamming window is applied to each of these blocks to minimize spectral leakage. 
Following \citet{stahl2021distinctions}, statistical symmetry across the jet centerline axis is exploited by mirroring and appending the snapshots to the data matrix, thus doubling the number of blocks used. 
These SPOD parameters are consistent across all cases.

\begin{figure}
    \centering
    \includegraphics[width=\textwidth,trim={10 25 10 20},clip]{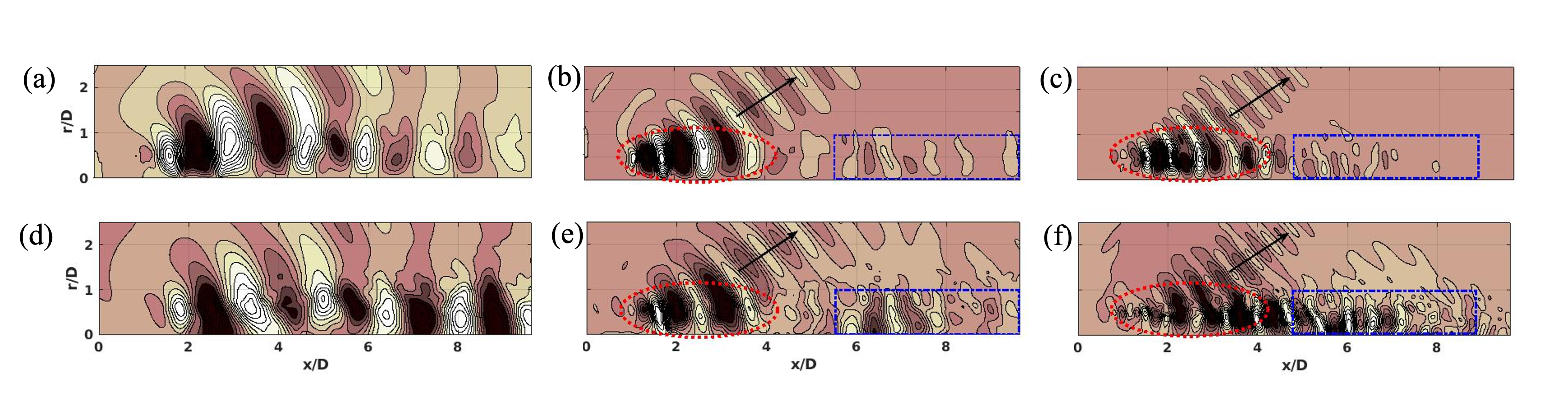}
    \caption{Leading SPOD Modes of $\partial \psi'_a/\partial x$ from LES (top) compared with those of $\Theta'$ obtained from schlieren (bottom) at \textit{St}=0.34 (left), \textit{St}=0.63 (middle) and \textit{St}=1.02 (right).}
    \label{fig:GE404SPODm1}
\end{figure}

Figure~\ref{fig:GE404SPODm1} compares the leading SPOD Modes of $\Theta'$ obtained from the schlieren images with those of $\partial \psi'_a/\partial x$, the pure acoustic component obtained from the LES, as a function of Strouhal number (\textit{St}$=f U_j/D$).
To mimic the procedure used when experimental schlieren is available, 
the $\Theta'$ values are computed using a single 2D plane, whereas the truth model values are
the $\partial \psi'_a/\partial x$ obtained by solving the Poisson equation for $\psi'_a$ (eqn.~\ref{Doak_irr2}) in the full 3D LES domain.

\begin{figure}
    \centering
    \includegraphics[width=\textwidth]{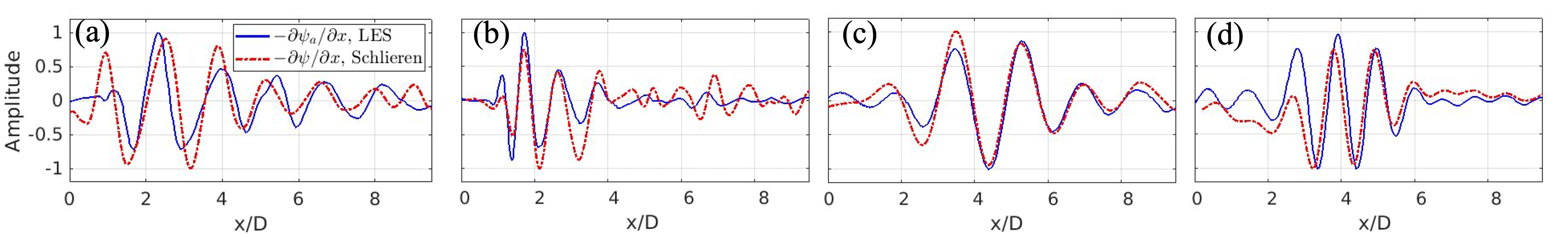}
    \caption{Comparison of leading $-\partial \psi'_a/\partial x$ SPOD Modes from LES with leading $\Theta'$ SPOD Modes from schlieren: (a) \textit{St}$=0.34$ at $r/D=0.5$, (b) \textit{St}$=0.63$ at $r/D=0.5$, (c) \textit{St}$=0.34$ at $r/D=2.0$, and (d) \textit{St}$=0.63$ at $r/D=2.0$.}
    \label{fig:lineplot}
\end{figure}
As expected, the leading SPOD modes from the LES consist of a Kelvin-Helmholtz (K-H) wavepacket~\citep{schmidt2018spectral} with a strong downstream noise radiation signature analogous to a supersonically travelling wavy-wall~\citep{Tam2009mach}.
With an increase in frequency, the wavelength of the radiation and the streamwise extent of the radiating region becomes shorter (highlighted with a red oval) and a secondary Orr wavepacket~\citep{schmidt2018spectral} appears downstream (marked with a blue rectangle).
These features are accurately captured by the leading SPOD Modes of $\Theta'$ extracted from the schlieren images. 
Figure~\ref{fig:lineplot} compares the streamwise variation of the SPOD modes at \textit{St}$=0.34$ and \textit{St}$=0.63$ along $r/D=0.5$ and $r/D=2.0$.
The schlieren derived $\Theta'$ accurately captures the peak and valleys associated with the acoustic wavepackets in the LES even along the jet lip-line.
This is a very useful observation since it shows that $\Theta'$, although containing contributions from both acoustic and thermal components, provides an accurate representation of the SPOD Modes of the pure acoustic component in a heated jet.

\subsection{Test Case 2: Unheated under-expanded Mach $1.27$ Impinging Jet}
This flow-field consists of a $D=0.0254$m circular nozzle exit, located at  $x/D=0.5$, exhausting perpendicular to a ground plate located $4D$ downstream.
This database has also been extensively validated against experimental measurements including near-field pressure and shadowgraphs in \cite{prasad2021exchange} and \cite{stahl2021distinctions}.

\begin{figure}
   \centering
      \includegraphics[width=0.7\textwidth,trim={0 0 250 0},clip]{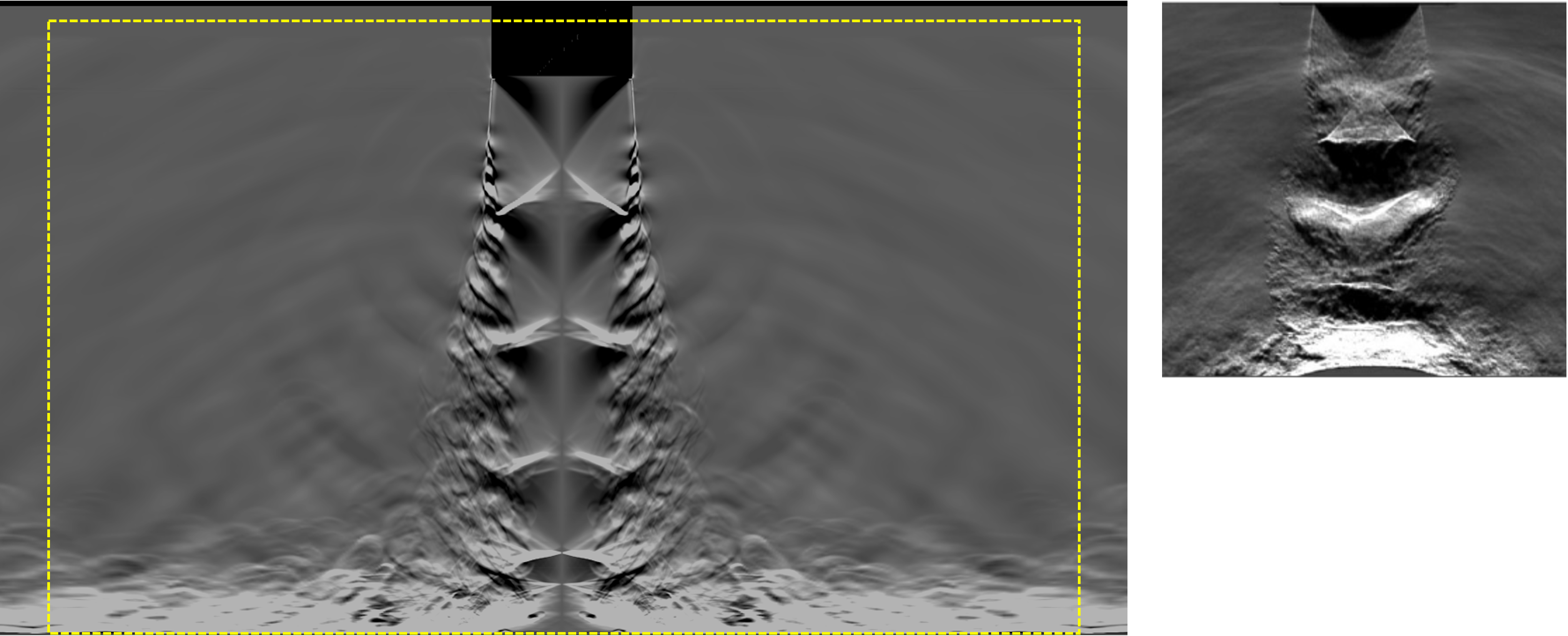}    
   \caption{Instantaneous Schlieren snapshot with the sponge zone location for the single impinging jet test case.}
   \label{fig:SIJSchlieren}
\end{figure}

Figure~\ref{fig:SIJSchlieren} shows an instantaneous schlieren image obtained from the 3D flow-field, extending $5D$ radially on either side of the jet centerline. 
The flow-field can be notionally divided into three regions that collectively constitute components of a feedback loop. 
The free jet region consists of a thin boundary layer emanating from the nozzle exit that rolls up into coherent structures due to the K-H instability. 
The turbulent structures impinge on the ground plate and are diverted along the wall jet region. 
The potential core of the free jet, that consists of shock-cells due to the under-expanded nature of the jet, is terminated by a stand-off shock near the wall marking the beginning of the impingement zone. 
The interaction of the coherent structures 
with the stand-off shock and the ground plate results in the generation of upstream travelling acoustic waves, which  
provide a periodic forcing of the thin shear layer at the nozzle exit.
The feedback process results in two high-intensity tones at \textit{St}~$=0.43$ and \textit{St}~$=0.55$~\citep{stahl2021distinctions}.

\begin{figure}
    \centering
    \includegraphics[width=0.8\textwidth]{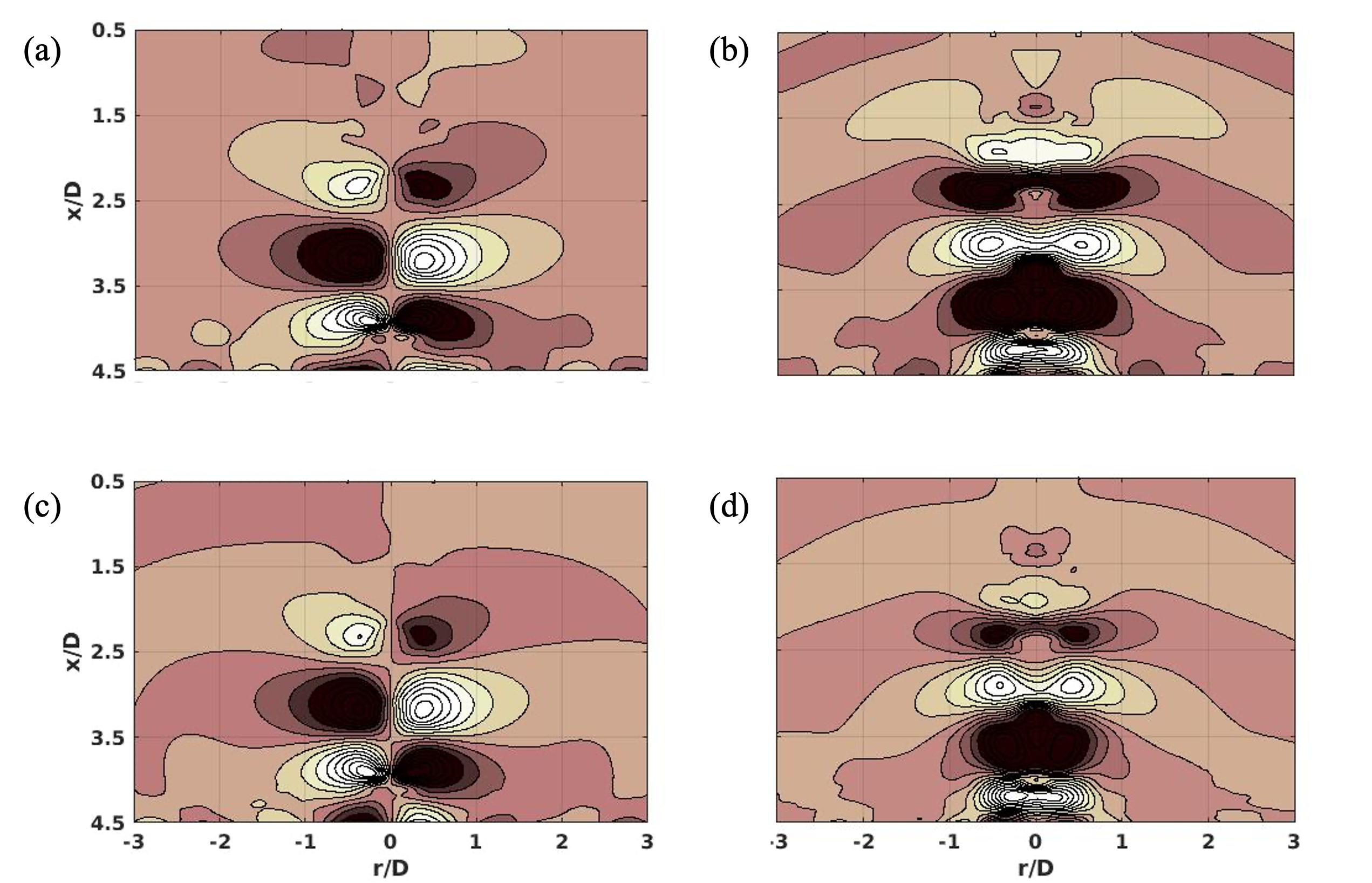}
    \caption{Leading SPOD Modes corresponding to the two dominant impingement tones from LES (top) and Schlieren (bottom): \textit{St}$=0.43$ (left) and \textit{St}$=0.55$ (right) . 
    }
    \label{fig:SIJSPOD}
\end{figure}

A sequence of $1{,}600$ schlieren images is generated at $54.1$kHz and subjected to the same procedure as above, with a sponge zone (Fig.~\ref{fig:SIJSchlieren}).
For the LES data, the MPT procedure is applied to the full 3D flow-field.
Figure~\ref{fig:SIJSPOD} compares the leading SPOD Modes of $\partial \psi'_a/\partial x$ from the LES with those of $\Theta'$ at the two feedback frequencies.
The \textit{St} $=0.43$ tone is associated with an asymmetric mode shape, as evident from the opposite color pairs of wavepackets on either side the jet centerline in Fig.~\ref{fig:SIJSPOD}(a), whereas
the \textit{St} $=0.55$ tone 
is comprised of an axisymmetric mode as shown in Fig.~\ref{fig:SIJSPOD}(b).
These mode shapes are accurately captured by the $\Theta'$ SPOD modes obtained from the schlieren as shown in Figs.~\ref{fig:SIJSPOD}(c) and~\ref{fig:SIJSPOD}(d) respectively.
This demonstrates the robust nature of the present method, as a change in test case from a free to an impinging jet, does not require any modifications to the solution procedure.

\subsection{Test Case 3: Unheated over-expanded Mach 1.35 Twin Rectangular Jet}
This TRJ configuration~\citep{ghassemi2021control} consists of two identical sharp-throat, military-style rectangular converging-diverging nozzles at an aspect ratio of two and a design Mach number ($M_d$) of $1.5$. 
The nozzle exit width ($w$) and height ($h$) are $24.13$ mm and $12.06$ mm respectively. 
This results in an area-based equivalent diameter of $D_e = 19.25$ mm. 
The center-to-center spacing between the nozzles is $2.25 D_e$. 

\begin{figure}
    \centering
    \includegraphics[width=0.9\textwidth,trim=0 0 0 0,clip]{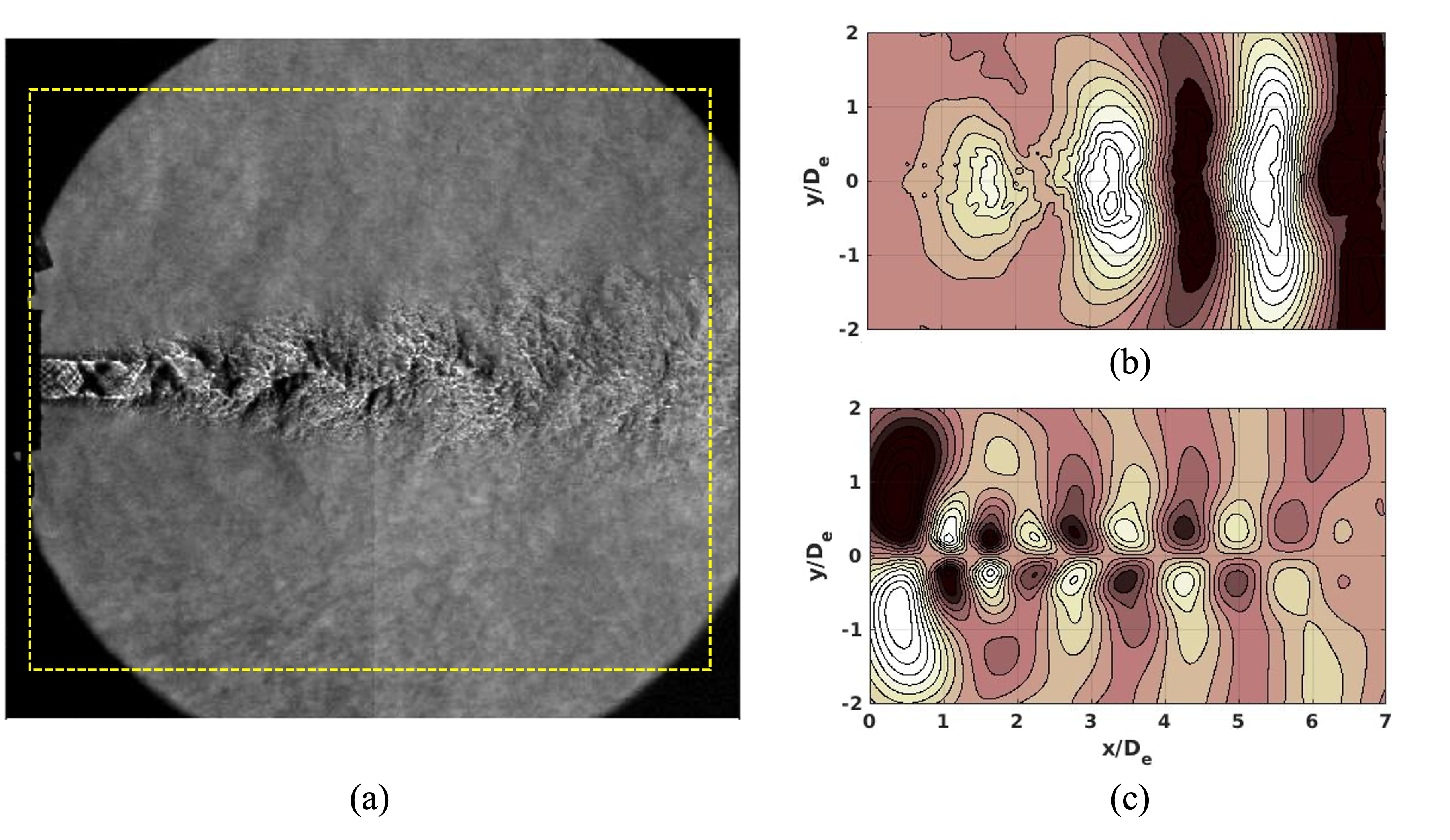} \\
    \caption{Instantaneous schlieren image along the minor axis for the TRJ test case (left), and leading SPOD Modes of $\Theta'$ at \textit{St}$=0.30$ (top right) and \textit{St}$=0.41$ (bottom right) extracted from a sequence of 2000 schlieren images.}
    \label{fig:TRJSchlieren}
\end{figure}

The flow-field is characterized by an in-phase coupling between the two jets resulting in a screech tone at \textit{St}$=0.41$ that dominates the minor axis (defined as the plane bisecting the longer edge of the nozzle). 
Figure~\ref{fig:TRJSchlieren}(a) shows an instantaneous schlieren image along the minor axis of the TRJ configuration obtained from a standard Z type schlieren system.
Since the jets are coupled in-phase, a field of view along the minor axis shows only one of the two jets.
The MPT procedure is applied to a sequence of $2{,}000$ such images sampled at $40$kHz in the same manner as before. 
The location of the sponge region is highlighted in Fig.~\ref{fig:TRJSchlieren}(a).
Figures~\ref{fig:TRJSchlieren}(b) and~\ref{fig:TRJSchlieren}(c) show the leading SPOD modes of $\Theta'$ at two frequencies corresponding to the peak shallow angle far-field noise (\textit{St}$=0.30$) and the screech frequency (\textit{St}$=0.41$) respectively.
As expected, with an increase in frequency, the wavepacket represented by $\Theta'$ is confined  closer to the jet exit.
This is consistent with known noise source characteristics of jets, where the higher frequencies are radiated from further upstream.
Additionally, the SPOD mode at \textit{St}$=0.30$ is symmetric about the jet axis and shows a clear downstream radiation pattern.
This is representative of the super-directive nature of the jets which dominates the peak noise radiation direction. 
The SPOD mode at \textit{St}$=0.41$ on the other hand, shows a flapping behavior as evident from the opposite color pairs of wavepackets on either side the jet centerline, and exhibits both upstream and downstream radiating waves. 
The former are indicative of the screech phenomenon at this frequency, whereas the out-of-phase nature of the wavepacket on either side of the jet centerline is %
consistent with the near-field measurements and coherence calculations at the same operating conditions~\citep{ghassemi2021control}.
This demonstrates the capability of the present method to accurately capture the different roles played by the acoustic waves in TRJ dynamics by imposing a physical constraint of irrotationality. 

\section{Conclusion \label{sec:Conclusion}}
A method is developed to extract acoustic content from time-resolved schlieren images through Doak's Momentum Potential Theory (MPT), which is not constrained by flow geometry or problem dependent parameters. 
Three test cases, including single free and impinging round jets and a free rectangular twin jet in imperfectly expanded condition, demonstrate the robustness of technique. 
Irrotational wavepacket structures are filtered from the schlieren images by solving a Poisson equation derived from MPT. 
When combined with SPOD, the resulting modes of the filtered irrotational component accurately capture the primary roles played by acoustic waves in the flow-field, irrespective of the configuration. 

The present technique has the potential to provide real-time tracking of the jet acoustic content in the near-field which can be leveraged for feedback control. 
The technique offers new avenues of simulation-experiment fusion 
analogous to~\citet{berry2017low}, who projected POD spatial basis functions from PIV data onto a corresponding LES flow-field to calculate time coefficients. 
A similar fusion between numerical and schlieren-derived MPT-derived wavepackets, such as for instance using the former to overcome the 2-D  nature of the latter could greatly expand the information derived from relatively sparser data.
Finally, although the test cases presented in this investigation are focused on jet dynamics, the procedure is also valid for other flow-fields of aerodynamic interest such as, airfoils and cavity flows, among others, where flow-acoustic interactions are crucial.

\section*{Acknowledgements}
This work was performed in part under the sponsorship of the Office of Naval Research with Dr. S. Martens serving as the Project Monitor. 
The views and conclusions contained herein are those of the authors and do not represent the opinion of the Office of Naval Research or the U.S. government. 
The authors are grateful for the use of the experimental data collected by the Ohio State research team led by Prof. M. Samimy. Contributors include Ata Esfahani and Dr. Nathan Webb. 

\clearpage
\bibliography{Ref}

\end{document}